\documentclass[11pt]{article}
\usepackage{moriond,epsfig}
\usepackage{graphicx}
\usepackage{capt-of}

\def\PLB{{\em Phys. Lett.}  B}
\def\PRL{\em Phys. Rev. Lett.}

\def\be{\begin{equation}}
\def\ee{\end{equation}}
\def\bea{\begin{eqnarray}}
\def\eea{\end{eqnarray}}

\def\met{$\not{\!\!\! E_T}$}
\def\mmet{\not{\!\! E_T}}
\def\pb{pb$^{-1} $ }
\def\d0{D{\O}}

\begin{document}
\vspace*{4cm}
\title{SEARCHES FOR SUPERSYMMETRY AT THE TEVATRON}

\author{ ELSE LYTKEN \\
	(for the CDF and \d0 collaborations)}

\address{Department of Physics, Purdue University \\ 525 Northwestern Avenue, West Lafayette, Indiana, USA}

\maketitle\abstracts{
The results for searches for Supersymmetry at the Tevatron Collider are summarized in this paper. We focus here on searches for chargino/neutralino and the lightest stop, as well as
scenarios with R-parity violation and split supersymmetry. 
No significant excesses with respect to the Standard Model were observed and constraints are set on the SUSY parameter space. 
}

\section{Introduction}
Supersymmetry (SUSY), an attractive extension of the Standard Model (SM), is based on a new symmetry between fermions and bosons. Each Standard Model particle acquires a supersymmetric partner with $\Delta spin = 1/2$. 
Most of the SUSY models considered assume the conservation of a quantum number, R-parity\footnote{$ R_P = (1)^{3(B-L) +2s}$ },
resulting in a stable lightest supersymmetric particle, the LSP. This stable, weakly interacting particle would escape detection and give rise to missing transverse energy, { \met}.
In minimal supergravity models the lightest supersymmetric particle is usually the lightest neutralino, $\tilde \chi^0_1$.
The non-excluded masses for the proposed superpartners are typically of the order of 100 GeV/c$^2$ and above, thus potentially accessible at the Tevatron. 
The two Tevatron experiments, \d0 and CDF have successfully been taking data at $\sqrt{s} = $1.96 TeV. Results shown are based on 300 - 800 \pb of data.

\section{Squarks and gluinos}
\subsection{Searching for stop}
Due to their strong couplings, squarks and gluinos are expected to be produced abundantly at hadron colliders. The lightest stop quark, $\tilde t_1$, is in many scenarios expected to be the lightest of
all the squarks, thus making it very favorable to search for at the
Tevatron. Recently a search for light stop quarks in
final states with $e^{\pm}\mu^{\mp}+b \bar b +\mmet$ was completed with the \d0 detector. This is the signature of $\tilde t_1$ pair production, followed by $\tilde t_1 \rightarrow b \ell \tilde \nu $, 
 which is expected to be the dominant decay mode when $m_{\tilde \nu} \sim m_W$. The analysis defines 3 regions, each optimized for a particular mass difference, $\Delta M = m_{\tilde t_1} - m_{\tilde \nu}$. The results are shown in Table~\ref{tab:stop} below.
\begin{table}[h]
\caption{\d0 results on the search for 
$\tilde t_1 \tilde{\bar t_1} \rightarrow e^{\pm}\mu^{\mp}+b \bar b +\mmet$ (350 \pb)\label{tab:stop} }
\begin{center}
\begin{tabular}{|c|c|c|}
\hline
  $ \Delta M$ & SM expectation & Observed\\ \hline
$ 20-40$ GeV/c$^2$    & $ 22.99\pm3.10 $ & 21 \\ 
$ 50-60$  GeV/c$^2$   & $ 34.63\pm3.96 $ & 34\\ 
$ \geq 70$  GeV/c$^2$ & $ 40.66\pm4.38$  & 42\\ \hline
\end{tabular}
\end{center}
\end{table}

\begin{figure}[h]
\begin{minipage}{7.2cm}
\begin{center}
\includegraphics[height=2in]{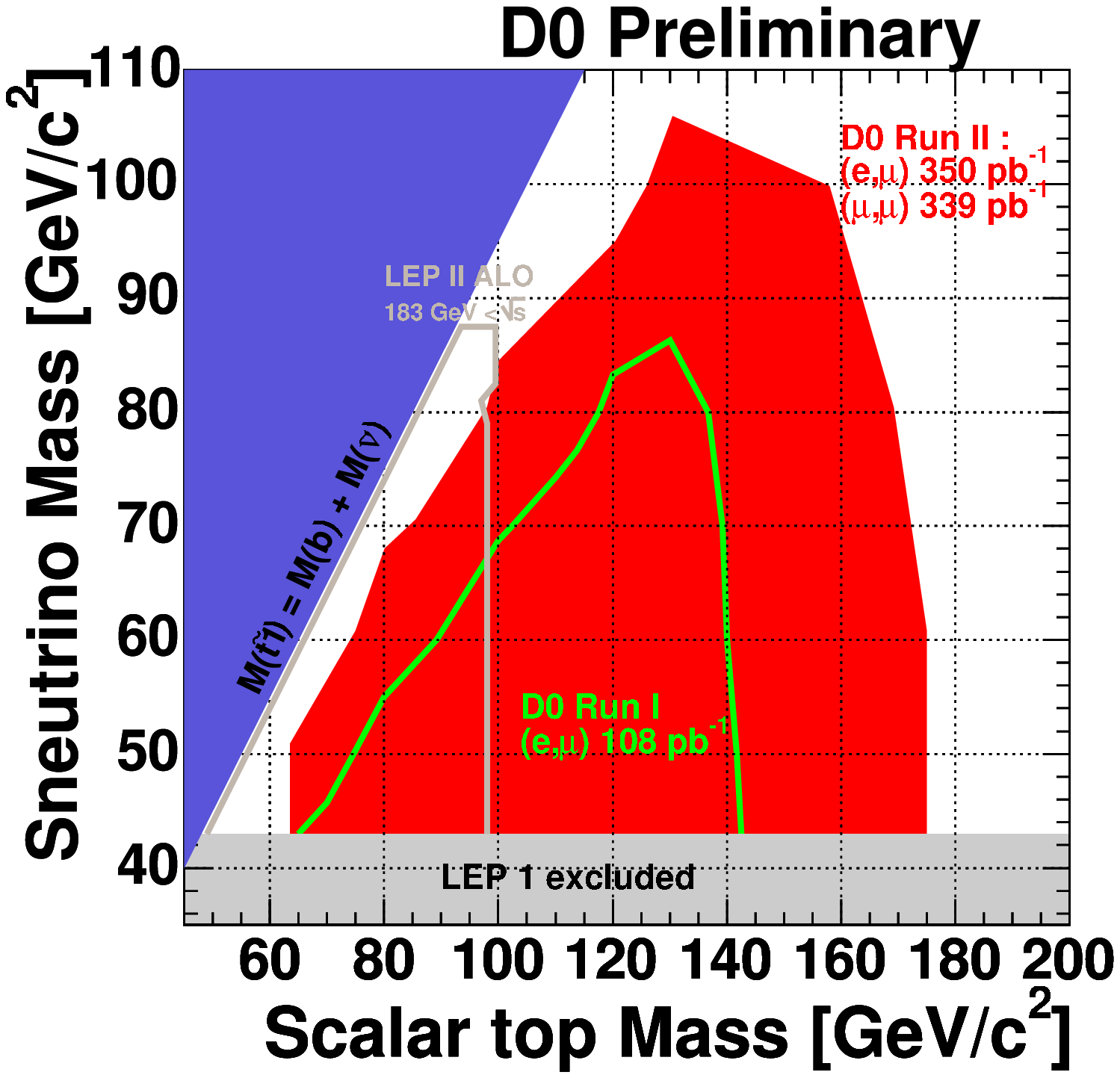}
 \caption{95\% CL exluded region in the stop search for the combination of $e\mu$ and $\mu\mu$ final states.}
\label{fig:squark} 
\end{center}
\end{minipage}
\hfill
\begin{minipage}{8cm}
\begin{center}
\includegraphics[height=1.7in]{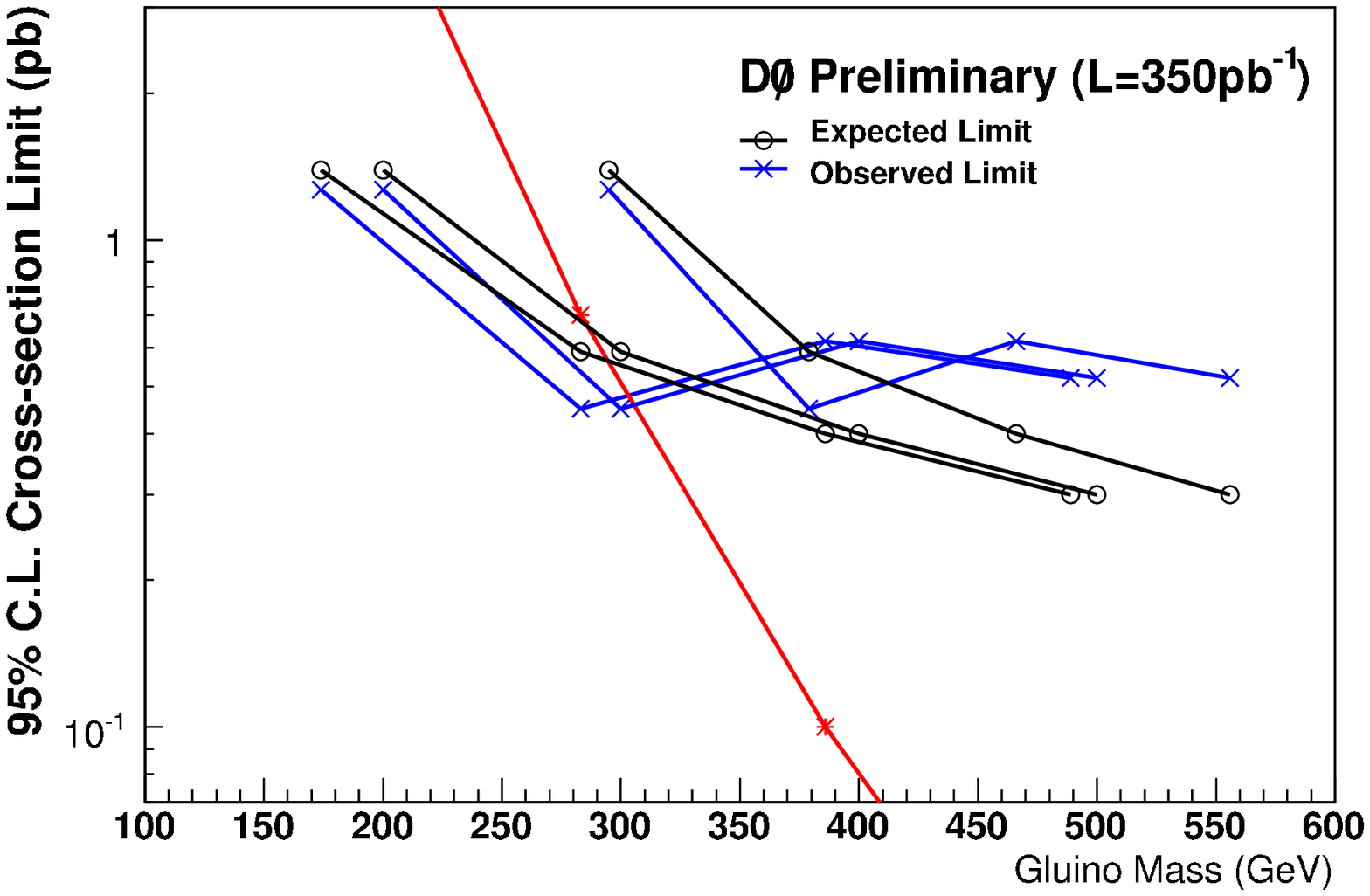} 
\caption{Stopped gluino search: 95\% CL limits for expected (blue crosses) and observed (black, open) for 3 masses: $m_{\tilde \chi^0_1}$ = 50, 90, 200 GeV/c$^2$. } \label{fig:gluino}
\end{center}
\end{minipage} 
\end{figure}

\noindent This result was combined with the previous result~\cite{ref:d0-stop}, the $\mu^{\pm}\mu^{\mp}+b \bar b +\mmet$ channel. No excess above the SM expectation was observed in any of the channels and the resulting combined limit is shown in figure~\ref{fig:squark}.

\subsection{Stopped gluinos}
\d0 has also looked into a possible signature for split supersymmetry~\cite{ref:splitsusy}, where the new SUSY scalars can be much heavier than the fermions. In that case the decay of the gluino into squarks would be heavily suppressed and the gluino therefore could have a lifetime long enough to stop in the calorimeters and decay to $g+\tilde{\chi}^0_1$. This would appear as an event with one high $E_T$ shower and thereby implicit large { \met}, out of time with a collision, due to the long lifetime of the gluino. The data used for the search is collected by a trigger for diffractive physics, and the energy spectrum of the shower is studied for signal and backgrounds. The analysis finds no hints of stopped gluinos in the measured spectrum and thus proceeds to set a limit on the gluino mass. Figure~\ref{fig:gluino} shows the resulting exclusion region.

\section{Chargino-neutralino}
\subsection{Multilepton final states}
The cross section for the associated production of $\tilde{\chi}^{\pm}_1 $ and $\tilde{\chi}^0_2$ is expected to be significant at the Tevatron due to the light masses. 
 In the case where both decay leptonically, $\tilde \chi^{\pm}_1 \rightarrow \ell^{\pm}\nu\tilde\chi^0_1$ and $\tilde \chi^0_1 \rightarrow \ell^{\pm}\ell^{\mp}\tilde\chi^0_1$, one can have very clean final states of 3 leptons, plus {\met} caused by the $\tilde{\chi}^0_1$'s and the neutrino. Very few SM processes contribute to such a signature, the dominant being Drell-Yan plus a misidentified jet or a $\gamma\rightarrow e^+e^-$ conversion, and real $WZ$ production. CDF presented new results on trileptons looking at many different decay channels. Results are summarized in Table~\ref{tab:trilep}. The { \met} spectrum for the $\mu\mu$ channel is illustrated in Figure~\ref{fig:trilep}. No significant excess is observed and limits are set on the $\chi^{\pm}_1$ mass in minimal models. A previous D{\O} analysis~\cite{ref:d0-cn} sets a limit of  $m_{\chi^{\pm}_1} \geq $117 GeV/c$^2$ for $m_{\tilde \ell} \sim m_{\tilde \chi^0_2}$.
\begin{table}[h]
\caption{CDF results of search for 
$\tilde \chi^{\pm}_1 \tilde \chi^0_2 \rightarrow \ell \ell + X$ ({\tiny $^*$= low $p_T$ triggers}) \label{tab:trilep} }
\begin{center}
\begin{tabular}{|c|c|c|c|c|c|c|} \hline
Channel & $ee + \ell$ & $ \mu\mu + \ell$  & $ \mu e + \ell$ & $ ^* ee + track$ & $ ^* \mu\mu + \ell$ & $\ell^{\pm} \ell^{\pm} $ \\ 
\small $\mathcal{L}$ & \small 346 \pb & \small 745 \pb & \small $\sim$ 700 \pb & \small  607 \pb & \small  312 \pb &  \small 704 \pb\\ \hline
SM expectation  & $ 0.17\pm 0.05$ & $ 0.64\pm 0.18$ & $ 0.78 \pm 0.11$ & $ 0.49\pm 0.10$ & $ 0.13\pm0.03 $ & $ 6.8\pm 1.0 $  \\
Observed & 0 & 1 & 0 & 1 & 0 & 9\\ \hline
\end{tabular}
\end{center}
\end{table}

\subsection{Diphotons}
\begin{figure}[h]
\begin{minipage}[h]{6.2cm}
\begin{center}
\includegraphics[width=\textwidth, height=0.8\textwidth]{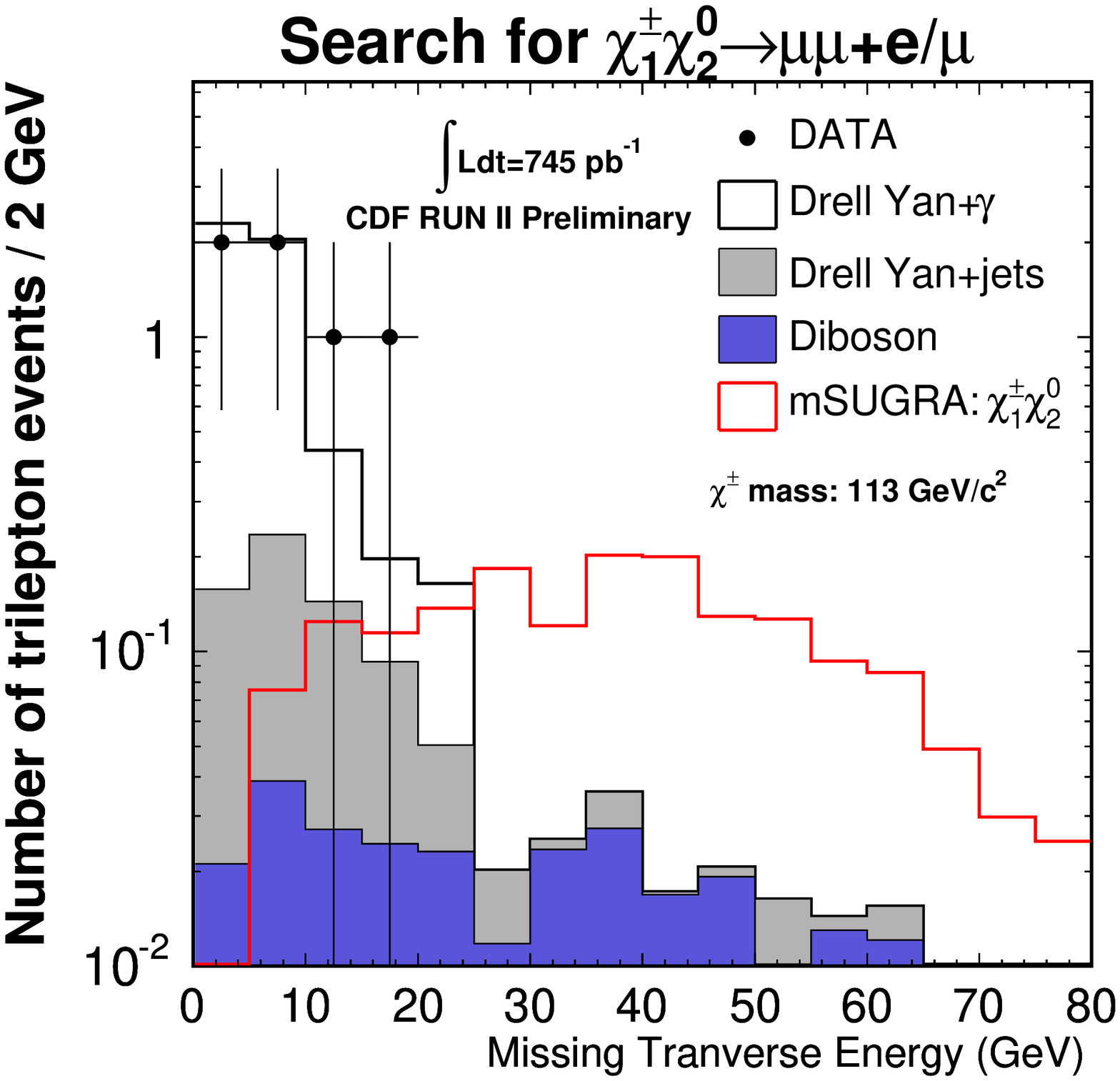} \caption{{\met} spectrum before the cut of the trilepton $\mu\mu + \ell$ channel.} \label{fig:trilep}
\end{center}
\end{minipage}
\hfill
\begin{minipage}[h]{6.9cm}
\begin{center}
\includegraphics[width=\textwidth]{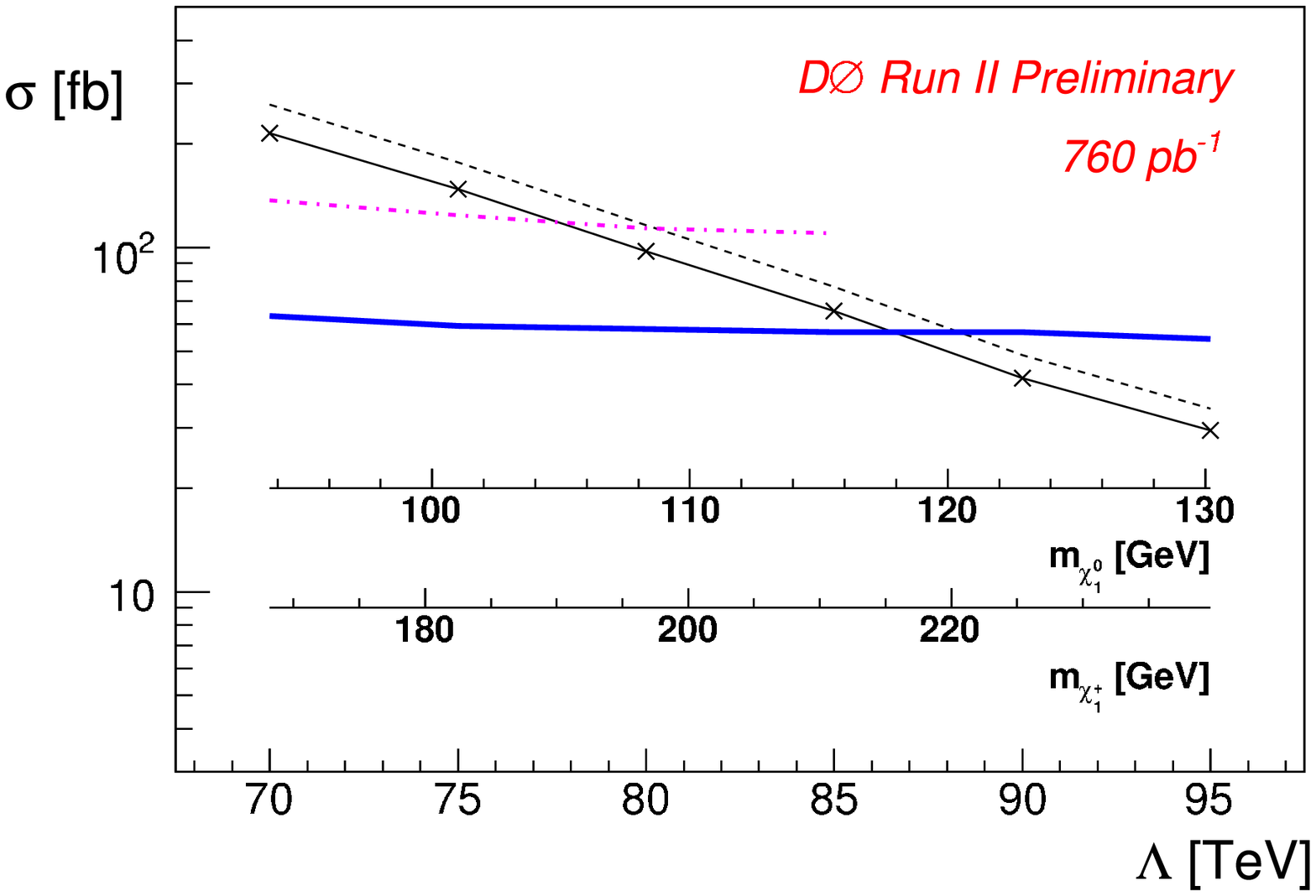} \caption{Results of search for GMSB $\tilde\chi\tilde\chi$ production. Full blue line is current result, pink dashed-dotted line is the previous result from \d0.}\label{fig:gmsb}
\end{center}
\end{minipage}
\end{figure}

In another popular SUSY model, Gauge Mediated Supersymmetry Breaking (GMSB), the LSP is the gravitino, the super partner of the graviton. Each of the $\tilde{\chi}^0_1$'s from chargino-neutralino production will decay into a gravitino and a photon. Independently of the decays of the charginos and other neutralinos, each pair produced set of gauginos will therefore have a final state with two photons and \met. D{\O} has searched for this signature in 760 \pb of data. Four events are observed with an expected background of 2.1$\pm$0.7 events.
The resulting limits are shown in Figure~\ref{fig:gmsb}.

\section{R-parity violation}
\subsection{$\tilde \chi^0_1 \rightarrow \mu\mu\nu$}
Imposing R$_P$ conservation is not a necessity in supersymmetry. The Tevatron experiments therefore also look for signatures that could arise from decays where R-parity is violated. R-parity violation
would add more terms to the supersymmetric Lagrangian, for instance for
lepton flavor violation: \be W_{\not \! R_P} = \lambda_{ijk} L_i L_j \bar E_k \ee
A non-zero $\lambda$ would cause decays of the lightest neutralino into neutrinos and charged leptons.
A CDF analysis looking for an excess of $eel(l)$ and $\mu\mu l(l)$ final states, with $l=e,\mu$, is sensitive to both $\lambda_{121}$ and  $\lambda_{122}$ couplings. The observation is compatible with expectation given the uncertainties in 346 \pb of data. Both the 3 and 4 lepton signatures were investigated and combined to set limits on the couplings: $\sigma<$0.21 pb for $\lambda_{121}> 0 $ and 0.11 pb for $ \lambda_{122}>0 $.

\subsection{Long-lived LSP}
\begin{figure}[ht]
\begin{minipage}[b]{0.45\columnwidth}%
    \centering
    \includegraphics[width=0.9\textwidth]{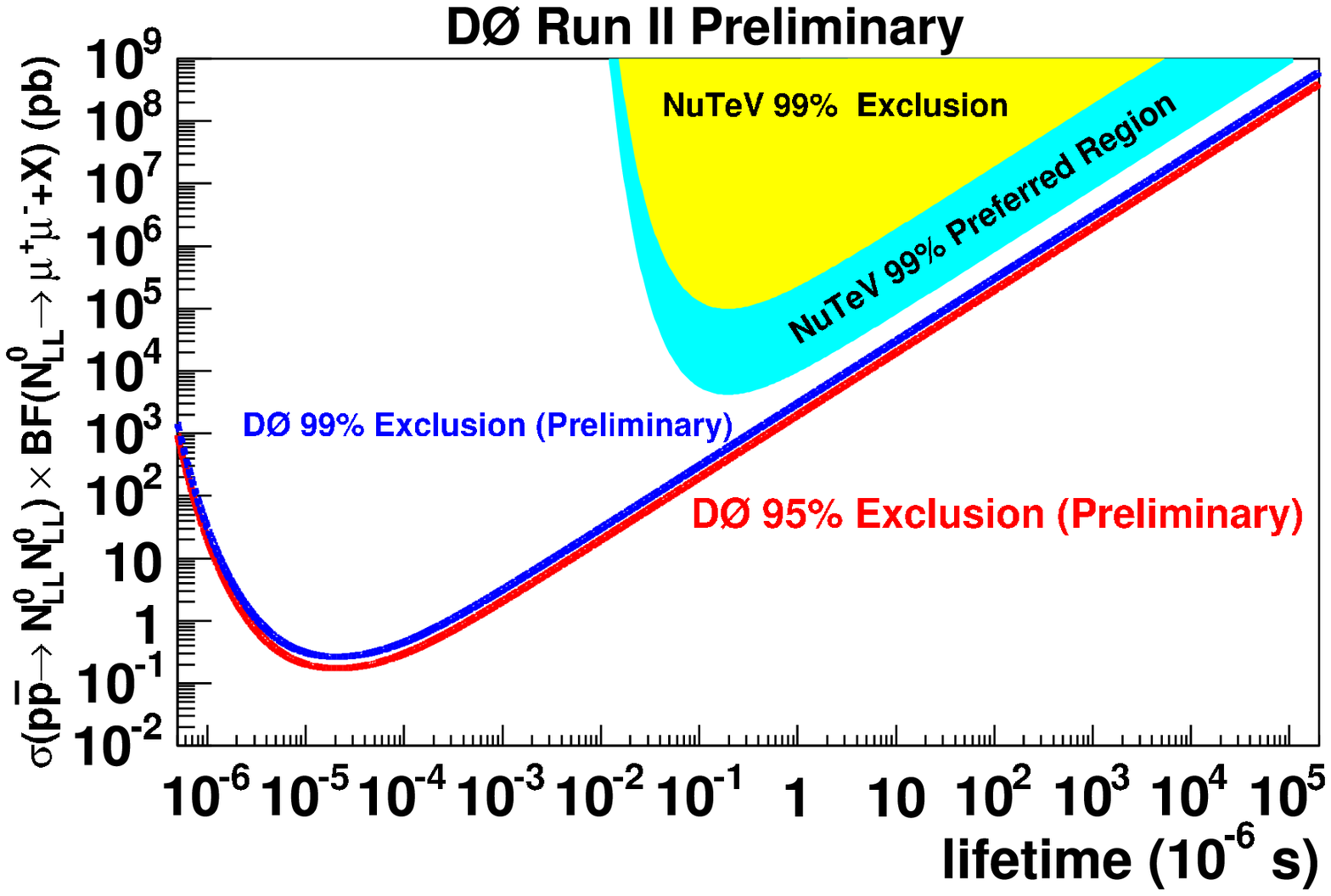}
    \caption{Red curve is the 95 \% CL limit on pair production of neutral, long-lived particles.}\label{fig:lllsp}
\end{minipage}%
\hfill%
\begin{minipage}[b]{0.48\columnwidth}%
\begin{tabular}{|c|c|c|c|c|}
\hline 
 & \small Expected & \small Obs & \small Expected & \small Obs  \\ \hline
\small $ B_s \rightarrow \mu\mu$ & \small $0.88 \pm 0.30$ & \small 1 & \small $0.39 \pm
 0.21$\small  &\small  0\\ 
\small $ B_d \rightarrow \mu\mu$ & \small $1.86 \pm 0.34$ & \small 2 & \small
$0.59 \pm 0.21$ & \small 0\\ \hline
\end{tabular}
\captionof{table}{CDF results from $B_s \rightarrow \mu\mu$ search. The left and right parts correspond to different muon selections.\label{tab:bsmumu} }
\end{minipage}
\end{figure}

\d0 looked for the decay of the neutralino to leptons and a neutrino in 383 \pb of data. The analysis focuses on the scenario where the $\not \!\! R_P$ coupling is weak and the LSP would travel $\geq$ 5 cm before decaying. This possibility was
inspired by an excess in dimuon events reported by NuTeV~\cite{ref:nutev}. No
events are observed with an expectation of 0.8$\pm$1.1$\pm$1.1 from
backgrounds and the limit set (Figure~\ref{fig:lllsp}) excludes the possibility that the NuTeV events
are due to neutralino decay.

\section{B$\rm _s$ to $\rm \mu^{\pm}\mu^{\mp}$}
In addition to the direct searches for supersymmetry the Tevatron experiments
are also trying to constrain supersymmetry by searching for rare decays, such as B$\rm _s \rightarrow
\mu\mu$. This decay is heavily suppressed in the SM but can be enhanced
orders of magnitude if new particle loops (like SUSY) exist. The
latest CDF analysis looks for opposite-sign $\mu\mu$ pairs in the B$\rm _s$ and B$\rm _d$ mass windows, using sidebands to estimate the expected backgrounds. Results with 780 \pb of data are shown in table~\ref{tab:bsmumu}. The observations are in agreement with background expectations, and the new limits, BR(B$\rm _s \rightarrow\mu\mu$)$<$1.0$\cdot$10$^{-7}$, and BR(B$\rm _d \rightarrow\mu\mu$)$<$3.0$\cdot$10$^{-8}$, improve the latest published result~\cite{ref:bsmumu} by a factor of 2.  
\section{Conclusions}
CDF and \d0 have searched for new physics in the form of supersymmetry and find no evidence for such new physics. The limits presented are the worlds best at this time. With the Tevatron and the experiments now working optimally we expect to chart much more uncovered territory in the coming few years.


\end{document}